\documentclass[
aps,nofootinbib,showpacs,showkeys,preprint
tightenlines
] {revtex4}

\usepackage{epsf,epsfig,subfigure,axodraw,graphicx,amsmath,amssymb}
\usepackage{color}

\newcommand{\five}{${\bf 5}$}
\newcommand{\fiveb}{${\bf\overline 5}$}
\newcommand{\ten}{${\bf 10}$}
\newcommand{\tenb}{$\overline{\bf 10}$}

 \def\LN{LN$\chi$}

\def\MGUT{$M_{\rm GUT}$}

\def\suf1{{SU(5)$\times$U(1)$_X$}\ }
\def\sut1{{SU(2)$\times$U(1)$_Y$}\ }

\def\fivebt{\overline{\bf 5}}
\def\fivet{{\bf 5}}
\def\tenbt{\overline{\bf 10}}
\def\tent{{\bf 10}}
\def\onet{{\bf 1}}

\begin{document}

\title{\Large\bf White dwarf axions, PAMELA data, and flipped-SU(5)}

\author{Kyu Jung Bae,$^{(a)}$ Ji-Haeng Huh\email{jhhuh@phya.snu.ac.kr},$^{(a)}$ Jihn E. Kim\email{jekim@ctp.snu.ac.kr},$^{(a)}$ Bumseok Kyae,$^{(a)}$ and Raoul D. Viollier$^{(b)}$}
\affiliation{ $^{(a)}$Department of Physics and Astronomy and Center for Theoretical Physics, Seoul National University, Seoul 151-747, Korea\\
$^{(b)}$Institute of Theoretical Physics and Astrophysics, Department of Physics, University of Cape Town, Private Bag, Rondebosch 7701,
South Africa
 }
\begin{abstract}
Recently, there are two hints arising from physics beyond the standard model. One is a possible energy loss mechanism due to emission of very weakly interacting light particles from white dwarf stars, with a coupling strength $\sim 0.7\times 10^{-13}$, and another is the high energy positrons observed by the PAMELA satellite experiment. We construct a supersymmetric flipped-SU(5) model, \suf1 with appropriate additional symmetries,  [U(1)$_H]_{\rm gauge}\times$[U(1)$_R\times$U(1)$ _\Gamma]_{\rm global}\times Z_2$, such that these are explained by a very light {\it electrophilic} axion of mass 0.5 meV from the spontaneously broken U(1)$_\Gamma$ and two component cold dark matters from $Z_2$ parity. We show that in the flipped-SU(5) there exists a basic mechanism for allowing excess positrons through the charged SU(5) singlet leptons, but not allowing anti-proton excess due to the absence of the SU(5) singlet quarks. We show the discovery potential of the charged SU(5) singlet $E$ at the LHC experiments by observing the electron and positron spectrum. With these symmetries, we also comment on the mass hierarchy between the top and bottom quarks.
\end{abstract}


\maketitle

\section{Introduction}

The most looked-for particles on Earth at the present time are axions \cite{KimCarosi} and weakly interacting massive particles(WIMPs) \cite{highEpartDM}. Recently, hints on these particles have been reported from outer space sources. The 10 year old INTEGRAL data \cite{INTEGRAL} gives the 511 keV line.  The white dwarf cooling has been suggested by the emission of very weakly interacting light particles(VWLP) (mass less than $\sim 1$ eV) with a very small coupling strength to electron ($\sim 0.7\times 10^{-13}$) \cite{Isern08}. At present, this white dwarf bound can be considered just as an upper bound, but in this paper we adopt a bold assumption that their best fit corresponds to the existence of a VWLP. More recently, the remarkable observation of high energy positron excess in the satellite PAMELA data \cite{PAMELAe} hints heavy (100 GeV -- 10 TeV) cold dark matter(CDM) particles \cite{
Bergstrom08,TwoDMs,Raidal08}. Since the 511 keV line can be explained by the astrophysical origin, or by the particles of mass ${\cal O}(1-10)~{\rm MeV}$ \cite{Partgammaray}, or by an excited state almost degenerate to the WIMP \cite{Weaner07}, in this paper we attempt to understand the latter two of these observations from a grand unification (GUT) viewpoint of the flipped-SU(5) \cite{Barr82,DKN84flip}.

The PAMELA group \cite{PAMELAe} reports the high energy positron excess, above 10 GeV up to 60 GeV, at the level  $e^+/(e^++e^-)\sim O(0.1)$, but has not reported any noticeable excess of anti-protons \cite{PAMELAp}. The charactersistic of the data between 10--60 GeV is a slightly rising positron flux, which is inconsistent to the lightest supersymmetric particle(LSP) in the minimal supersymmetric standard model(MSSM) with supersymmetric(SUSY) particles at ${\cal O}(100~\rm GeV)$ \cite{TwoDMs}. Thus, the MSSM with ${\cal O}(100~\rm GeV)$  lightest neutralino DM component $\chi$(\LN) has to be modified if SUSY has to be kept for the gauge hierarchy solution. The minimal extension needs just three two component fermions $N_R, E_R$ and $E_R^c$ \cite{TwoDMs}, with the coupling
\begin{equation}
W=e_RE_R^cN_R.\label{eEcN}
\end{equation}
If one tries to introduce more fields around TeV scale, certainly the rising positron flux can be explained, but it is very difficult \cite{Raidal08} to explain nonobservation of antiproton flux. This suggests a different treatment of electron from quarks. Indeed, the flipped-SU(5) GUT \suf1 treats charged leptons differently from quarks \cite{Barr82} such that the charged lepton is an \suf1 singlet,
\begin{widetext}
\begin{equation}
\tent_1=\left(\begin{array}{ccccc}
0& d& -d& u^c& -d^c\\ -d& 0& d& -u^c& d^c\\
d& -d& 0& u^c& -d^c\\
-u^c& u^c& -u^c& 0& \nu_0\\
d^c& -d^c& d^c& -\nu_0& 0
\end{array}
\right)_R,\
\fivebt_{-3}=\left(\begin{array}{c}
u\\ u\\ u\\  \overline{e}\\ -\overline{\nu}_e\end{array}\right)_R,\
\onet_{5}=e_R~ .\label{flipRep}
\end{equation}
\end{widetext}

The specific form of the coupling (\ref{eEcN}) is related to the problems of the Yukawa couplings. In this regard, we note the five old but fundamental problems on the masses of the standard model fermions. Firstly, there is a hierarchy of the top quark mass $m_t\sim fv/\sqrt2\simeq 170$ GeV at the electroweak scale of $v\simeq 250$ GeV, and the rest of the standard model particle masses are much smaller than the top quark mass, $m\le m_b\simeq 4.5~{\rm GeV}\ll m_t$. Second, there is a sizable Cabibbo mixing, especially between the first and the second family quarks. Third, the quark mass ratio is reversed in the first family compared to those of the second and the third family quarks, i.e. ${m_u}/{m_d}<1$ while ${m_c}/{m_s}>1$ and ${m_t}/{m_b}>1$. Even though it is not proper to define an inverted mass pattern for leptons, it looks like that the electron mass is also an inverted pattern because ${m_e}/{m_\mu}\simeq {m_u}/{4m_c}\sim {\cal O}(1/200)$. Fourth, there is a big hierarchy of singlet neutrino mass of $m_{\nu_0}\sim {\cal O}(10^{14})$ GeV for the seesaw mechanism \cite{Minkowski77} and the electroweak scale of $v\simeq 250$ GeV. Finally, we point out that the SM neutrino mass pattern shows a tri-bi maximal mixing pattern \cite{tribiPerkins}. Except the tri-bi maximal mixing, in this paper we try to understand the remaining four problems in a flipped-SU(5) GUT.

To have a very light axion implied by the white dwarf data, we introduce a Peccei-Quinn (PQ) symmetry U(1)$_\Gamma$ so that the symmetry we consider is extended to \suf1 times U(1)$_\Gamma$. To have the inverted mass pattern of the quark mass ratio in the first family, we assign different $\Gamma$ charges for the first family members and the second and the third family members. But, to have a large Cabibbo mixing, we require Q$_{\rm em}=\frac23$ quarks or Q$_{\rm em}=-\frac13$ quarks mix fully, restricted only by the strength of the corresponding Yukawa couplings but without any restriction from the symmetry arguments. In the flipped-SU(5), since both the quark singlet $d_R$ and the quark doublet $q_R^c$ appear in ${\bf 10}_1$, viz. Eq. (\ref{flipRep}), it is easy to mix Q$_{\rm em}=-\frac13$ quarks fully without restrictions from the symmetry arguments. So, we assign the same symmetry charges for three members ${\bf 10}_1^{(i)}(i=1,2,3)$. Then, Q$_{\rm em}=-\frac13$ quarks can be mixed fully as desired. But the charges of $\overline{5}_{-3}^{(i)}$ and ${1}_{5}^{(i)}$ can be different for different $i$ to allow the inverted mass pattern of the quark mass ratio for the first family. For the singlet neutrino mass problem, we need a large VEV, above the axion window of $10^9-10^{12}$ GeV.  If there is no other gauge symmetry, the PQ symmetry may be broken by this large VEV. Thus, we introduce another U(1) gauge symmetry which will be called a horizontal gauge symmetry U(1)$_H$, distinguishing the first family and the second and third families. Finally, to keep $E_R-E_R^c$ pair at the electroweak scale and to have proton stability and two dark matter components \cite{TwoDMs}, we introduce the $R$ symmetry and $Z_2$ symmetry (the $R$-parity). Thus, the symmetry we consider is
\begin{equation}
SU(5)\times U(1)_X\times U(1)_H\times U(1)_R\times U(1)_{\Gamma}\times Z_2.\label{symmetry}
\end{equation}

In this model, the renormalizable couplings are gauge couplings and the top quark Yukawa coupling. All the other Yukawa couplings are non-renormalizable. The model we introduce does not have the $c_3$ term of Ref. \cite{KimCarosi}, where the $c_3$ term is the axion and QCD anomaly coupling, and the PQ charges of Higgs doublets vanish. Therefore, the very light axion we obtain is not a proper Kim-Shifman-Vainstein-Zakharov (KSVZ) axion \cite{Kim79},  or the Dine-Fischler-Srednicki-Zhitnitskii (DFSZ) axion \cite{DFSZ81}, but can be called a `variant very light axion'. The electron--axion coupling is known to be small for the KSVZ axion, as calculated in Ref. \cite{Sredecoupling85}. In our case, the axion arises from the $c_2$ term of Ref. \cite{KimCarosi}, where the $c_2$ term is the light quark mass term, through a non-renormalizable interaction, and the electron--axion coupling turns out to be large. Since this `very light variant axion' is required to have a large electron coupling, it will be called a very light {\it electrophilic} axion.

In Sec. \ref{sec:WD}, we briefly review the white dwarf bound \cite{Isern08}. In Sec. \ref{sec:flip}, we present a flipped-SU(5) model, satisfying the criteria presented above. In Sec. \ref{sec:phenomenology}, we discuss the SUSY phenomenology for the model presented in Sec. \ref{sec:flip}. Sec. \ref{sec:conclusion} is a conclusion.

\section{White Dwarf Bound}\label{sec:WD}
White dwarfs can give us hints about their last stage evolution. White dwarfs, with the mass range less than $1.08 M_\odot$, are the fossil stars after consuming all their nuclear fuel. Here the electron degeneracy pressure is the one supporting the gravitational collapse, and the surface temperature ranges down from ten times the solar surface temperature. The temperature difference between the core and the surface is not greater than 100 K, and hence if the VWLP is produced at the core then its mass cannot be much greater than 1 eV. Since the white dwarf has a relatively simple structure and its evolution mechanism is just based on the cooling, its luminosity function has been rather accurately predicted. Thus, if there is a novel energy loss mechanism by a VWLP like axion, it can affect the resulting luminosity function of the white dwarf which can be used to constrain the properties of that particle or give a clue for its existence.

The most strong bound on VWLP comes from the existence of the sharp cutoff at the faint part in the luminosity function ($n$) versus the bolometric magnitude ($M_{\rm bol}$) plot. The sharp cutoff is due to the finite age of our Galaxy.
The luminosity function is given by \cite{Isern08}
\begin{equation}
  n(M_{\rm bol})=\int^{M_u}_{M_l}\quad \phi(M)\psi(T_G-t_{\rm cool}-t_{\rm ps})\tau_{\rm cool}dM,
\end{equation}
where $\phi(M)$ is the initial mass function, $\psi$ is the birth rate of the progenitor star, $M_{\rm bol}$ is the bolometric magnitude, $M$ is the mass of the progenitor star and $\tau_{\rm cool}=dt/dM_{\rm bol}$ is the characteristic time scale of the white dwarf cooling. Although the luminosity function depends on the birth rate which is not yet exactly known, the bright part of the curve is insensitive to it up to the overall normalization. Thus, one can exploit the bright part to constrain the hypothetical cooling mechanism due to the VWLP.

In this regard, recently the possible energy loss rate has been computed \cite{Isern08}. If a very light boson with mass less than 1 eV is produced inside white dwarfs, its coupling to electron is estimated as
$
0.7\times 10^{-13}.
$
Any boson with this coupling strength to electron can explain the white dwarf evolution {\it a la} Ref. \cite{Isern08}, using Ref. \cite{Catalan08}.

Since axions are well-motivated light pseudoscalar bosons, Ref. \cite{Isern08} tries to interpret the coupling $0.7\times 10^{-13}$ in terms of  the DFSZ model \cite{DFSZ81}.
In the interior of the white dwarf, the axion bremstralung coming from the axion-electron Yukawa coupling, which is significant in the DFSZ model but absent in the hadronic axion model, is dominant over the Primakoff process. The energy loss rate per mass with the crystalization effect is given by \cite{Isern08},
\begin{equation}
  \varepsilon=1.08\times10^{23}\frac{g_{aee}^2}{4\pi}
  \left( \frac{Z^2}{A} \right)\left( \frac{T}{10^7{\rm K}} \right)^4 F(\rho,T)\quad {\rm erg~ g^{-1}s^{-1}},
\end{equation}
where $F(\rho,T)$ is the correction factor due to the crystalization effect and $g_{aee}$ is the Yukawa coupling between the electron and the axion. In the DFSZ model, it corresponds to $2.8\times10^{-11} \sin^2\beta (m_a/1{\rm eV})$ with the domain wall number 6. From this study, Isern et al. \cite{Isern08} point out that the small deviations of the white dwarf luminosity function from the standard cooling model can constrain the axion mass $m_a\sin^2\beta < 10$ meV, and even presented the best fit of the luminocity function with the axion emission for $m_a\sin^2\beta \simeq 5$ meV. In the DFSZ model, this white dwarf bound on the axion mass corresponds to the axion decay constant $F_a> 0.6\times 10^9\sin^2\beta$ GeV and the best fit corresponds to $F_a\simeq 1.2\times 10^9\sin^2\beta$ GeV. Below, we try to construct a flipped-SU(5) to accomodate the best fit of Ref. \cite{Isern08}.

The electron axion coupling can be written as $(m_e\Gamma(e)/F)\bar e i\gamma_5 ea$, where $F=N_{DW}F_a$ and $\Gamma(e)$ is the PQ charge of $e$. Since the axion mass is given by $0.6~{\rm eV}(10^7~{\rm GeV})/F_a$, the best fit of Isern et al. \cite{Isern08}  in the DFSZ model interpreted as the electron--axion coupling in any axion model,
\begin{widetext}
\begin{equation}
{\rm Best~fit:}\quad \left|\frac{m_e\Gamma(e)}{F_a}\right|=\frac{m_e}{0.72\times 10^{10}~\rm GeV}\simeq 0.7\times 10^{-13},\quad {\rm in~ any~ axion~ model. }
\label{aeWDcoupling}
\end{equation}
\end{widetext}
If the domain wall number is large and the PQ charge of electron is 1, then we need to have a smaller $F_a$ than $10^{10}$ GeV since $N_{DW}F_a$ is required to be $0.72\times 10^{10}~\rm GeV$. Generally, the domain wall number is counted as $\frac12$ for any quark  carrying the unit PQ charge. Therefore, if only one Dirac quark carries the PQ charge 1 so that each chirality contributes one unit of the PQ charge, then the domain wall number is 1. In this way \cite{Isern08}, the DFSZ model gives $N_{DW}=6$ since there are six quarks, and the electron--axion coupling is $m_e/(6\times 1.2\times 10^9~{\rm GeV})$ with $F_a=1.2\times 10^{9}$ GeV.

There is an interesting case that only one Weyl quark carries the PQ charge 1. In this case, our formula for the axion--electron coupling must use $N_{DW}=\frac12$.\footnote{There cannot be a half integer domain wall number. It simply means that with a proper definition, the electron PQ charge is 2 compared to the quark charge 1. See Eq. (\ref{aecoupling}).\label{footDW}} In such a model Eq. (\ref{aeWDcoupling}) implies $F_a\simeq 1.4\times 10^{10}$ GeV. Below we concentrate on the latter possibility in a flipped SU(5) model, obtaining Eq. (\ref{aeWDcoupling}) for the very light {\it electrophilic} axion. This moves $F_a$ to the supergravity favored scale \cite{Kim84common} even with the large electron--axion coupling envisioned in Ref. \cite{Isern08}.

\section{A flipped SU(5) model}\label{sec:flip}

SO(10) GUT groups cannot realize our idea of Ref. \cite{TwoDMs}. The SO(10) GUT with the SU(5) GUT assignment \cite{GG73} does not introduce our SU(5) singlet $E$ and $E^c$ fields since the electromagnetic charge operator $Q_{\rm em}$ is fully embedded in SU(5)$_{\rm GG}$.  The flipped-SU(5) assignment of hypercharge in the SO(10) subgroup \cite{Barr82},  namely ${\bf 16}={\bf 10}_{1} +\overline{\bf 5}_{-3} +{\bf 1}_5$, allows the SU(5) singlet (in ${\bf 1}_5$) carrying the electromagnetic charge, and $e_R$ does not belong to ${\bf 10}_{1}$ or $\overline{\bf 5}_{-3}$. Supersymmetric (SUSY) generalization of the flipped-SU(5) chain from SO(10), including the vacuum expectation value of Higgs field not belonging to the origin of the weight diagram to reduce the rank of the gauge group, has been discussed in \cite{DKN84flip}. A simple choice of such symmetry breaking with ${\bf 10}_H$ and $\overline{\bf 10}_H$ from string construction was shown to be possible in the fermionic construction \cite{Ellis87} and in the orbifold compactification \cite{KimKyaeflip}. Also, in the SUSY flipped-SU(5), the doublet-triplet splitting problem is solved elegantly as shown in the fermionic construction \cite{Ellis89} and in the orbifold compactification \cite{KimKyaeflip}.

Here, we do not consider the SO(10) breaking chain to an \suf1 gauge group. The \suf1 gauge group is considered as an ultraviolet completion gauge group, possibly arising from a compactification of string theory. As in Ref. \cite{TwoDMs}, we use the right-handed chiral fields so that the electron singlet is represented as $e_R$ instead of $e_L^c$ and the quark doublet is represented as $q^c_R$ instead of $q_L$. This right-handed chiral field notation is simpler to discuss the interaction $e_RE_R^c N_R$ which we try to introduce for the PAMELA data.

We use the following right-handed notation for representations of the flipped-SU(5). The electroweak hypercharge is
\begin{equation}
Y=\frac15 Y_5 -\frac{1}{5}X\label{Yflipped}
\end{equation}
where the SM singlet generator embedded in SU(5) is
\begin{equation}
Y_5 = {\rm diag.}\left(\frac{-1}{3}~
\frac{-1}{3}~\frac{-1}{3}~~\frac{1}{2}~~\frac{1}{2}
\right).
\end{equation}
The representations (\ref{flipRep}) are consistent with this definition.

The matter field ${\bf 10}_1$ contains the quarks doublet $({\bf 3^*,2})_{-1/6}$, the quark singlet $({\bf 3,1})_{-1/3}$ and the neutral lepton singlet $({\bf 1,1})_{0}$. $\overline{\bf 5}_{-3}$ contains the quark singlet $({\bf 3,1})_{2/3}$ and the anti-lepton doublet $({\bf 1,2})_{1/2}$. $({\bf 1,1})_{5}$ is the charged lepton of the unit charge --1. Therefore, it is useful to remember that singlets $d_R$, $u_R$ and $e_R$ are members of ${\bf 10}_1$, $\overline{\bf 5}_{-3}$ and $({\bf 1,1})_{5}$, respectively.

\subsection{Breaking flipped SU(5)}

The singlet neutral direction ($\nu_0$ and $\bar\nu_0$ directions) of ${\bf 10}_1$ and $\overline{\bf 10}_{-1}$ can be assigned to GUT scale vacuum expectation values(VEVs), breaking SU(5)$\times$U(1)$_X$ down to SU(3)$\times$
SU(2)$\times$U(1)$_Y$. For the spontaneous symmetry breaking of the flipped SU(5), thus, we need a vectorlike pair, ${\bf 10}_H$ and $\overline{\bf 10}_{H}$, which contain the SM singlet components, $\nu_0$ and $\overline{\nu}_0$.  The following superpotential,
\begin{eqnarray}
W_H=Z({\bf 10}_H\overline{\bf 10}_H-M_G^2)
\end{eqnarray}
allow ${\bf 10}_H$ and $\overline{\bf 10}_H$ to develop VEVs along the direction of $\nu_0$ and $\overline{\nu}_0$. Here the superfield $Z$ is a singlet of the flipped SU(5). Since the D flat direction is along $\langle {\bf 10}_H\rangle=\langle
\overline{\bf 10}_H\rangle$, thus, at the SUSY minimum we have
\begin{eqnarray}
\langle {\bf 10}_H\rangle=\langle \overline{\bf 10}_H\rangle =M_G.
\end{eqnarray}
The phase modes of $\overline{q}_H$ (SU(2) doublet) and $q_H$ (SU(2) doublet) in ${\bf 10}_H$ and
$\overline{\bf 10}_H$ correspond to the Nambu-Goldstone bosons and the real modes of $\overline{q}_H$ and $q_H$ become heavy as the Higgs boson becomes heavy in the Higgs mechanism in the standard model. The Nambu-Goldstone bosons are absorbed by the heavy gauge sectors when the flipped-SU(5) is broken. However, $d_H$ (SU(2) singlet) and $\overline{d}_H$ (SU(2) singlet) in ${\bf 10}_H$ and $\overline{\bf 10}_H$ need to be made somehow heavy. The quantum numbers of $Z$, ${\bf 10}_H$, $\overline{\bf 10}_H$, and some other superfields needed later are displayed in Table \ref{tab:Vectorlike}.
%
%
\begin{widetext}
\begin{table}[!h]
\begin{center}
\begin{tabular}{c|cccccc|cccc|cccc}
{\rm Fields}\ \ &\quad $Z$ &\quad $\Phi$ &\quad ${\bf 10}_H$ &\quad $\overline{\bf 10}_H$ &\quad ${\bf 5}_h$ &\quad $\overline{\bf 5}_h$ \quad  &\quad ${\bf 10}'$ &\quad
$\overline{\bf 10}'$ &\quad ${\bf 1}'$ &\quad $\overline{\bf 1}'$ &\quad $\xi$ &\quad $\overline{\xi}$ &\quad $\zeta$ &\quad $\overline{\zeta}$
\\
\hline
 $X$\ \ & \quad $0$ &\quad $0$ & \quad $1$ &\quad $-1$
& \quad $-2$ &\quad $2$ &\quad $1$ &\quad $-1$&\quad $5$ &\quad $-5$ &\quad $1$ &\quad $1$&\quad $1$ &\quad $1$
 \\
$H$\ \ & \quad $0$ &\quad $1$ & \quad $0$ &\quad $0$ & \quad $-1$ &\quad $-1$
& \quad $0$ &\quad $0$ &\quad $0$ &\quad $0$ &\quad $3$ &\quad $-3$&\quad $1$ &\quad $-1$\\
$R$\ \ & \quad $2$ &\quad $0$ & \quad $0$ &\quad $0$& \quad $2$ &
\quad $2$ & \quad $0$ &\quad $2$ &\quad $0$ &\quad $2$ &\quad $-8$ &\quad $0$&\quad $-8$ &\quad $0$\\
 $Z_2$\ \ & \quad $+$ &\quad $+$ & \quad $+$ &\quad $+$& \quad $+$ &
\quad $+$ &\quad $-$ &\quad $-$ &\quad $-$&\quad $-$
&\quad $-$ &\quad $-$&\quad $-$ &\quad $-$ \\
 $\Gamma$\ \ & \quad $0$ &\quad $0$ & \quad $0$ &\quad $0$ & \quad $0$ &\quad $0$
&\quad $1$ &\quad $-1$ &\quad $1$ &\quad $-1$&\quad $0$ &\quad $0$&\quad $0$ &\quad $0$ \\
\end{tabular}
\end{center}\caption{Vectorlike fields including the Higgs fields.  $Z$ and $\Phi$ are the flipped-SU(5) singlet fields. }\label{tab:Vectorlike}
\end{table}
\end{widetext}

The representation ${\bf 5}_{-2}$ contains $Q_{\rm em}=\frac13$ quark $\overline{D}$ and Higgs doublet ${H}_d$ with $Y=\frac{1}{2}$. The representation $\overline{\bf 5}_{2}$ contains $Q_{\rm em}=-\frac13$ quark ${D}$ and Higgs doublet $\overline{H}_u$ with $Y=-\frac12$. In order to decouple the modes
$\overline{D}$, ${D}$ in ${\bf 5}_{-2}$, $\overline{\bf 5}_{2}$, and also $d_H$, $\overline{d}_H$ in ${\bf 10}_H$, $\overline{\bf 10}_H$ from the low energy field spectrum, we consider the following superpotential,
\begin{eqnarray}
W_{D/T}=\frac{\Phi}{M_P}({\bf 10}_H{\bf 10}_H{\bf 5}_h
+\overline{\bf 10}_H\overline{\bf 10}_H\overline{\bf 5}_h ) .
\end{eqnarray}
We assume that the singlet $\Phi$, which carries the charge of the gauged U(1)$_H$, get a VEV of order $M_G$. We will provide later an explanation on how $\Phi$ can get the VEV. Because of the VEVs $\langle {\bf 10}_H\rangle$ and $\langle \overline{\bf 10}_H\rangle$ in the $\nu_0$ and $\overline{\nu}_0$ directions, $\overline{D}$ in ${\bf 5}_{-2}$ and $d_H$ in ${\bf 10}_H$, and also ${D}$ in $\overline{\bf 5}_{2}$ and
$\overline{d}_H$ in $\overline{\bf 10}_H$ obtain superheavy Dirac masses \cite{Ellis89,KimKyaeflip}.

\subsection{Yukawa couplings and family hierarchy}


The top quark mass is close to the \sut1 breaking scale and hence $t$ is required to obtain mass by the renormalizable terms. Since the $b$ and $\tau$ masses are O($10^{-2}$) of the top quark mass, it is reasonable to require for them to appear by nonrenormalizable terms. Thus, we allow the renormalizable superpotential term for the top quark and nonrenormalizable terms for the bottom quark and the tau lepton masses due to the gauge, global and discrete symmetries we introduce. For the second family members, the third family pattern is repeated but we allow ${\cal O}(10^{-2})$ couplings compared to the third family couplings. However, we treat the first family quite differently from the second and third families because the mass pattern is inverted for the up quark, i.e. ${m_u}/{m_d}<1$ while ${m_c}/{m_s}>1$ and ${m_t}/{m_b}>1$. We interpret the electron mass also by the inverted pattern because ${m_e}/{m_\mu}\simeq {m_u}/{4m_c}\sim {\cal O}(1/200)$. We investigate this kind of mass patterns by introducing the PQ symmetry U(1)$_\Gamma$ and the horizontal gauge symmetry U(1)$_H$.

As explained above, the $\overline{5}_{-3}$ representation of Eq. (\ref{flipRep}) in the flipped SU(5) accommodates a quark singlet $u_R$ ($Q_{\rm em}=\frac23$) [instead of $d_R$ as in SU(5)$_{\rm GG}$] and a lepton doublet $\overline{l}_R$ ($Q_{\rm em}=\frac12$). As a
result, considering only the \suf1 symmetry, $u$ type quarks and Dirac neutrino Yukawa couplings can be obtained from ${\bf 10}_{1}\overline{\bf 5}_{-3}\overline{\bf 5}_h$ with the dictated relation $m_u^T=m_{D\nu}$ in the flipped-SU(5) [rather than $m_d^T=m_{e}$ as in the SU(5)$_{\rm GG}$]. Since the singlet lepton $e_R$ remains also as a singlet of the flipped-SU(5) model unlike in the SU(5)$_{\rm GG}$, the Yukawa couplings for the charged leptons is given by ${\bf 1}_{5}\overline{\bf 5}_{-3}{\bf 5}_h$. The $d$ type quarks' Yukawa couplings come from the couplings of the type ${\bf 10}_{1}{\bf 10}_{1}{\bf 5}_h$ in the flipped-SU(5).

%
\begin{table}[!h]
\begin{center}
\begin{tabular}{c|ccc|ccc|c}
{\rm Fields}\ \ &\quad ${\bf 10}^{(2,3)}$ &\quad $\overline{\bf
5}^{(2,3)}$ &\quad ${\bf 1}^{(2,3)}$& ${\bf 10}^{(1)}$ &\quad $\overline{\bf 5}^{(1)}$ &\quad ${\bf 1}^{(1)}$ \quad &\quad $S$ \quad \\
\hline
 $X$\ \  &\quad $1$ &\quad $-3$ &\quad $5$\quad
 &\quad $1$ & \quad $-3$ &\quad $5$ &\quad $0$ \\
$H$\ \ &\quad $0$ &\quad $1$ &\quad $-1$ \quad & \quad $0$ &\quad $0$ & \quad $0$ &\quad $0$ \\
$R$\ \ &\quad $0$ &\quad $0$ &\quad $0$
 \quad & \quad $0$ &\quad $0$ & \quad $0$ &\quad $0$ \\
 $Z_2$\ \  &\quad $-$ &\quad $-$ &\quad $-$
\quad &\quad $-$ & \quad $-$ &\quad $-$ &\quad $+$ \\
 $\Gamma$\ \  &\quad $0$ &\quad $0$ &\quad $0$ \quad
 &\quad $0$ &\quad $-1$& \quad $-1$ &\quad $1$ \\
 \end{tabular}
\end{center}\caption{The matter fields and $S$. $S$ is an SU(5) singlet field whose VEV is breaking the PQ symmetry spontaneously.
}\label{tab:Families}
\end{table}
%
Therefore, we assign the Abelian charges given in Tables \ref{tab:Vectorlike} and \ref{tab:Families}.  The $Z_2$ symmetry distinguishes the matter ${\bf 10}^{(2,3)}$ from the GUT breaking Higgs ${\bf 10}_H$. With the U(1)$_H$ charges of Tables \ref{tab:Vectorlike} and \ref{tab:Families}, there is no U(1)$_H$--SU(5)--SU(5) anomaly. Introducing exotic singlets with $Q_{\rm em}=-\frac15$, $\xi, \overline{\xi}, \zeta$ and $\overline{\zeta}$ in Table \ref{tab:Vectorlike}, the mixed anomalies of U(1)$_H$--U(1)$_X^2$ and U(1)$_H^2$--U(1)$_X$ are made to vanish. A possible Tr$X$ anomaly is made to vanish by introducing the $H=0$ and $X= -1$ singlets in Table \ref{tab:H-Higgs}.

The Yukawa couplings for the second and third families of the MSSM matter fields, consistent with the symmetries of Tables \ref{tab:Vectorlike} and \ref{tab:Families}, are
\begin{widetext}
\begin{eqnarray} \label{3rd}
W^{(2,3)}=\frac{\Phi}{M_P}{\bf 10}^{(2,3)}{\bf 10}^{(2,3)}{\bf
5}_h + {\bf 10}^{(2,3)}\overline{\bf 5}^{(2,3)}\overline{\bf 5}_h
+ \frac{\Phi}{M_P}{\bf 1}^{(2,3)}\overline{\bf 5}^{(2,3)}{\bf 5}_h.
\end{eqnarray}
\end{widetext}
For the third family, therefore we qualitatively explain why the $b$ quark mass is much smaller than the $t$ quark mass, depending on the factor $\langle \Phi\rangle/M_P$.
The mystery of the quark masses is not in the mystery of the heaviness of top quark, but resides in the smallness of the masses of the first family members compared to the electroweak scale. Therefore, for the first family, we assign quantum numbers differently from those of the third family members. One simple choice is that the $d$ quark mass is still made small by the choice of Yukawa couplings but the up quark and electron masses arise from different type of couplings. This choice seems reasonable since the up quark mass is somehow inverted from the
second and the third generation pattern, $m_{2/3}>m_{-1/3}$. The electron mass is much smaller than the up quark mass and hence can be thought as the inverted pattern. Therefore, let us choose the quantum numbers of the first family members as shown in Table \ref{tab:Families}.

In  Table \ref{tab:Families}, we also show the singlet $S$ whose VEV breaks the PQ symmetry. The heavy fields ${\bf 10}', \overline{\bf 10}', {\bf 1}'$ and $\overline{\bf 1}'$ of Table \ref{tab:Vectorlike} will be integrated over to obtain the needed non-renormalizable interactions.
Note that for the $d$ type quarks the three flavors $\{d_R, s_R, b_R\}$ and $\{d_R^c, s_R^c, b_R^c\}$ can be fully mixed among themselves, since all the quantum numbers of ${\bf 10}^{(1)}$ are the same as those of ${\bf 10}^{(2,3)}$. This full mixing between $Q_{\rm em}=-1/3$ quarks enables one to generate a large Cabibbo angle, even though $Q_{\rm em}=2/3$ quarks are not mixed fully.

We assume that the VEV of the singlet $S$ is around the intermediate scale $10^{10}$ GeV. We will explain later how $S$ develops such a VEV.  The allowed nonrenormalizable Yukawa couplings $(\langle S\rangle/M_P){\bf 10}^{(1)}\overline{\bf 5}^{(1)}\overline{\bf 5}_h$ and $(\langle S\Phi\rangle/M_P^2){\bf 1}^{(1)} \overline{\bf 5}^{(1)}{\bf 5}_h$ by the gravity mediation gives, therefore, too small $u$ quark and electron masses. Thus, we introduce vectorlike
pairs of $\{{\bf 10}', \overline{\bf 10}'\}$ and $\{{\bf 1}', \overline{\bf 1}'\}$ with proper quantum numbers given in Table \ref{tab:Vectorlike}, and their heavy mass terms. Then, the Yukawa couplings relevant to the $u$ quark mass are
\begin{eqnarray}
W^{(u)}=\Phi{\bf 10}^{(1)}\overline{\bf 10}' + M'{\bf
10}'\overline{\bf 10}' + {\bf 10}'\overline{\bf
5}^{(1)}\overline{\bf 5}_h ,
\end{eqnarray}
where the mass parameter $M'$ can be smaller than $M_P$.  After integrating out the heavy $\{{\bf 10}',\overline{\bf 10}'\}$ pair, one can obtain the $u$ quark Yukawa coupling with a desired size,
\begin{eqnarray}
W^{(u)}=\frac{\langle \Phi\rangle}{M'}{\bf 10}^{(1)}\overline{\bf
5}^{(1)}\overline{\bf 5}_h ,\label{umass}
\end{eqnarray}
For the first family up-type quark, we have an extra factor $\langle \Phi\rangle/M'$ compared to the second and third family up-type quarks. So, with an ${\cal O}(10^{-1})$ smaller Yukawa coupling compared to the charm quark Yukawa coupling and $\langle \Phi\rangle/M'\sim {\cal O}(10^{-2})$, $m_u$ can be made smaller than $m_c$ by a factor  ${\cal O}(10^{-3})$.

Similarly, we have the following interactions for electron,
\begin{eqnarray}
W^{(e)}=S{\bf 1}^{(1)}\overline{\bf 1}' + M''{\bf 1}'\overline{\bf
1}' + \frac{\Phi}{M_P}{\bf 1}'\overline{\bf 5}^{(1)}{\bf 5}_h.
\end{eqnarray}
After integrating out ${\bf 1}'$ and $\overline{\bf 1}'$, the effective electron Yukawa coupling becomes
\begin{eqnarray}
W^{(e)}=-\frac{\langle S\Phi\rangle}{M''M_P}{\bf
1}^{(1)}\overline{\bf 5}^{(1)}{\bf 5}_h ,\label{eYukawa}
\end{eqnarray}
with $\langle S\Phi\rangle/M''M_P\sim 10^{-6}$.

This model does not have the $c_3$ term of Ref. \cite{KimCarosi} and the PQ charges of Higgs doublets vanish. Therefore, the very light axion we obtain is not a proper KSVZ or DFSZ axion. Out of  ${\bf 10}',\overline{\bf 10}'$, and  $\overline{\bf 5}^{(1)}$, carrying PQ charges, disregard the vectorlike ${\bf 10}'$ and $\overline{\bf 10}'$. Then, there remains a half quark in $\overline{\bf 5}^{(1)}$ carrying the PQ charge since the other half in ${\bf 10}^{(1)}$ does not carry the PQ charge. After integrating out heavy fields, this model therefore has the $c_2$ term of Ref. \cite{KimCarosi} and is a {\it variant very light axion} as some electroweak scale axions were called variant axions \cite{varaxion}. The mass term related to $c_2$ is a non-renormalizable interaction, and the electron--axion coupling will be shown to be large. The resulting $c_2$ type coupling is for the up quark mass term, $m_u\bar u_Lu_R e^{ia/F}+{\rm h.c.}$ \cite{KimCarosi}. Expanding $S$ in terms of the axion field, $S=(F+\rho)e^{2ia/F}$ with $\Gamma(S)=2$, Eq. (\ref{eYukawa}) gives the following axion--electron coupling,
\begin{align}
&-\frac{\langle |S|\Phi H_d\rangle}{M''M_P}
\left[e^{2ia/F} \bar e_L e_R+{\rm h.c.}\right]\nonumber\\
&= -m_e\bar ee +\left(m_e\frac{2}{F}\right) \bar ei\gamma_5 e~a\label{aecoupling}
\end{align}
In this model $F_a=F$ since effectively only one Weyl quark in $\overline{5}^{(1)}$ carries one unit of the PQ charge. This model gives effectively $N_{DW}=1/2$ since only a half quark carries the PQ charge. See footnote \ref{footDW}. Note that a more general electrophilic axion can result with the PQ charge carrying heavy quarks giving $c_3=N_{\rm h.q.}$ and $c_2=-c_3\pm\frac12$.

\subsection{Masses of singlet neutrinos}

For the seesaw generation of neutrino masses, it is necessary to have (i) the Dirac masses between the SM neutrinos and the singlet neutrinos at the electroweak scale and (ii) the heavy Majorana masses of ${\cal O}(10^{14})$ GeV for singlet neutrinos.
In string compactification to the MSSM, these two conditions are achievable in $Z_3, Z_{12-I}$ and $Z_{6-II}$ orbifold compactificaton \cite{KimSeesaw}.

In the presence of the U(1)$_R$ symmetry in the flipped-SU(5), it is not straightforward to write down the Majorana neutrino mass term of ${\cal O}(10^{14}~\rm GeV)$ for the singlet neutrino because it is embedded in ${\bf 10}_1$. So, the $R$ symmetry must be broken nontrivially by the VEVs of $R$-charge carrying scalar fields \cite{KS}.  We consider the following interactions:
\begin{eqnarray}
W^{(\nu)}=\frac{1}{M_P}{\bf 10}^{(i)}\overline{\bf 10}_H\Phi\Psi +
\frac{1}{M_P}\Phi\Phi'\Psi^2  ,
\end{eqnarray}
where the quantum numbers of $\Phi$, $\Phi'$, and $\Psi$ are shown in Table \ref{tab:H-Higgs}. We introduced $\Phi''$ in Table \ref{tab:H-Higgs} to make Tr$H=0$.
\begin{table}[!h]
\begin{center}
\begin{tabular}{c|ccccc|cc}
{\rm Fields}\ \ &\quad $4\eta$ &\quad ($\Phi$) &\quad $\Phi'$&\quad $\Phi''$ &\quad $\Psi$ \quad
&\quad ($S$) &\quad $\overline{S}$ \\
\hline
 $X$\ \ & \quad $-1$ & \quad $0$ &\quad $0$ &\quad $0$ & \quad $0$& \quad $0$ &\quad $0$\\
 $H$\ \ & \quad $0$& \quad $1$ &\quad $1$ &\quad $1$ & \quad $-1$& \quad $0$ &\quad $0$\\
 $R$\ \ & \quad $2$ & \quad $0$ &\quad $-2$ &\quad $0$ & \quad $2$& \quad $0$ &\quad $1$\\
 $Z_2$\ \ & \quad $-$& \quad $+$ &\quad $+$&\quad $-$ & \quad $-$& \quad $+$ & \quad $+$\\
 $\Gamma$\ \ & \quad $0$& \quad $0$ &\quad $0$&\quad $-1$ &\quad $0$&\quad $1$& \quad $-1$\\
\end{tabular}
\end{center}\caption{The $Q_{\rm em}=\frac15$ singlet $\eta$, and neutral singlets carrying nonzero U(1)$_H$ or PQ charges. We need four $\eta$s and one $\Phi''$ fields to make Tr$X=0$ and Tr$H=0$. Fields in the brackets are already listed in Tables \ref{tab:Vectorlike} and \ref{tab:Families}.}\label{tab:H-Higgs}
\end{table}
%

We assume that both $\Phi$ and $\Phi'$ get VEVs of order \MGUT. Then, $\Psi$ achieves a heavy mass. After integrating out $\Psi$, one can get the effective Majorana neutrino mass term,
\begin{eqnarray}
W^{(\nu)} = \frac{\langle \Phi\rangle\langle\overline{\bf
10}_H^2\rangle}{\langle \Phi'\rangle M_P} {\bf 10}^{(i)}{\bf
10}^{(j)}  .
\end{eqnarray}
Hence, if $\langle \Phi\rangle\sim\langle \Phi'\rangle$ and the dimensionless Yukawa coupling is of order unity, we get a Majorana neutrino mass of ${\cal O}(10^{14~}\rm GeV)$ as desired.

$\overline{\xi}$, $\overline{\zeta}$ and two of $\eta$s can achieve masses by $\langle \Phi\rangle$. To give masses to $\{\xi,\zeta, \eta_{(3)}, \eta_{(4)}\}$, we need more fields which we omit here. We note that these neutral singlets can be almost massless since their decoupling temperature is expected to be very large.

\subsection{The PQ symmetry breaking scale}
Now let us discuss how $\Phi$ and $S$ can achieve the VEVs. Tables \ref{tab:Vectorlike}, \ref{tab:Families}  and \ref{tab:H-Higgs} list the fields carrying nonzero U(1)$_H$ charges. The relevant D term potential of U(1)$_H$ is
\begin{eqnarray}
V_H=\frac{g_H^2}{2}\left||\Phi|^2+|\Phi'|^2
-|\Psi|^2 -\xi\right|^2  ,
\end{eqnarray}
where $\xi$ ($\sim M_G^2$) denotes the Fayet-Iliopoulos term \cite{FI}, and we have neglected the small contributions,  $|{5}_h|^2,|\overline{5}_h|^2,$ etc. Considering the soft mass terms, the minimum of the potential would be on $|\Phi|^2+|\Phi'|^2=\xi$. In addition,
consideration on the higher order term in the K${\rm
\ddot{a}}$hler potential, $K\supset
\frac{1}{M_P^2}\left|\Phi\right|^2\left|\Phi'\right|^2$, and its supergravity effect on the potential would determine the minimum at $\langle \Phi\rangle=\langle\Phi'\rangle=\sqrt{\xi/2}$. (In
supergravity, the Fayet-Iliopoulos term needs to be replaced by a proper modulus \cite{FIsugra}.)

Now, $S$ can develop the intermediate scale VEV ($\sim 10^{10}$ GeV) by the supergravity couplings
and the softly broken SUSY effects,
\begin{eqnarray}
W_{PQ}=\frac{1}{M_P}S^2\overline{S}^2  .\label{NonMGUT}
\end{eqnarray}
The soft SUSY breaking mass terms, the A-term corresponding to Eq. (\ref{NonMGUT}), and the F term contribution by Eq. (\ref{NonMGUT}) can make it possible to develop their VEVs  at the minimum of the scalar potential,
\begin{align}
&\langle S\rangle \sim \langle \overline{S}\rangle \sim
\sqrt{m_{3/2}M_P}\nonumber\\
 &\sim 10^{10}~{\rm GeV} \left(\frac{m_{3/2}}{10^{2~}\rm GeV}\cdot\frac{M_P}{10^{18~}\rm GeV}\right)^{1/2}.
\end{align}
At this scale, the PQ symmetry is completely broken. The gravitino mass of ${\cal O}(10^{2~}\rm GeV)$ leads to the axion scale of ${\cal O}(10^{10~}\rm GeV)$ we presented in Sec. \ref{sec:WD}.

\subsection{Two dark matter components}

In Ref. \cite{TwoDMs}, two DM components are introduced. In fact, there are two possible ultraviolet completions (items {\bf a} and {\bf b} below) with $N_R$ of Eq. (\ref{eEcN}). These are two high energy completion from the \LN\ and $N_R$ of of our model. These are based on how $N_R$ obtains mass.
\begin{itemize}
\item[\bf a.]
The needed matter fields in this case are $SU(3)\times SU(2)$ singlets $E_R, E_R^c, N_R$, and $N_R^c$. In the flipped-SU(5), they can appear as SU(5) singlets because the weak hypercharge depends on the SU(5) singlet generator $X$ also, as shown in Eq. (\ref{Yflipped}). The needed couplings are \cite{TwoDMs}
\begin{equation}
W_{DM}=e_RE_R^c N_R + N_R^3 ,\label{supotentialN3}
\end{equation}
where $e_R={\bf 1}^{(1)}_{5}$,  and hence we assign the charges shown in Table \ref{tab:PAMELAa}.
%
\begin{table}[!h]
\begin{center}
\begin{tabular}{c|cccc}
{\rm Fields}\ \ &\quad \quad $E_R^c$ &\quad $E_R$ &\quad $N_R$  &\quad $N_R^c$ ~
\\
\hline
$X$\ \ & \quad $-5$ & \quad $5$& \quad $0$& \quad $0$  \\
$H$\ \ & \quad $0$ & \quad $0$& \quad $0$& \quad $0$  \\
 $R$\ \ & \quad $\frac43$ & \quad $\frac83$
& \quad $\frac23$& \quad $\frac{10}3$\\
 $Z_2$\ \ & \quad $-$ & \quad $-$ &\quad $+$&\quad $+$\\
 $\Gamma$\ \ & \quad $1$ & \quad $-1$ &\quad $0$ &\quad $0$\\
\end{tabular}
\end{center}\caption{The SU(5) singlets for the coupling (\ref{eEcN}).
They do not carry U(1)$_H$ charges. }\label{tab:PAMELAa}
\end{table}
%
The $R$ symmetry forbids superheavy mass terms of $E_RE_R^c$ and $N_R^2$ as well as $\overline{H}_u{H}_d$.  Their mass term would be possible through the Giudice-Masiero mechanism
\cite{GMmech}:
\begin{eqnarray}
W\supset m_{3/2}E_RE_R^c+m_{3/2}'N_R N_R^c +
m_{3/2}''\overline{H}_u{H}_d ,
\end{eqnarray}
where $m_{3/2}$, $m_{3/2}'$, and $m_{3/2}''$ are of order the gravitino mass scale. The masses of $E$ and $N$ are at the O(100~GeV) scale.

\item[\bf b.] There is another method to give mass to $N_R$. It is by the $R$ charge assignment of $N_R$ be $\frac43$ so that the $N_R^3$ term of (\ref{supotentialN3}) is not allowed. The relevant quantum numbers are given in Table \ref{tab:PAMELAb}.
%
\begin{table}[!h]
\begin{center}
\begin{tabular}{c|ccc|c}
{\rm Fields}\ \ &\quad \quad $E_R^c$ &\quad $E_R$ &\quad $N_R$~~  &\quad $S'$ ~
\\
\hline
$X$\ \ & \quad $-5$ & \quad $5$& \quad $0$~~& \quad $0$  \\
$H$\ \ & \quad $0$ & \quad $0$& \quad $0$~~& \quad $0$  \\
 $R$\ \ & \quad $0$ & \quad $4$
& \quad $2$~~& \quad $4$\\
 $Z_2$\ \ & \quad $+$ & \quad $+$ &\quad $-$~~&\quad $+$\\
 $\Gamma$\ \ & \quad $1$ & \quad $-1$ &\quad $0$~~ &\quad $0$\\
\end{tabular}
\end{center}\caption{The SU(5) singlets for the coupling (\ref{eEcN}). $S'$ is superheavy. }\label{tab:PAMELAb}
\end{table}
%
 In this model, the following supergravity effect generates the following mass terms,
\begin{equation}
\int d^4\theta \frac{S^{\prime *}}{M_P}\left(\lambda E_R E_R^c+\lambda'N_R^2+\lambda'' \overline{H}_uH_d \right).
\end{equation}
At low energy, there are three more two component fields, $E_R, E_R^c,$ and $N_R$.
\end{itemize}
The $R$ charges of {\bf a} are different from those of Ref. \cite{TwoDMs} because in the present flipped-SU(5) model the $R$ charge of $\overline{H}_u H_d$ is necessarily nonvanishing while in  \cite{TwoDMs} it is zero. In the present case, the number of fields carrying the $R$ symmetry breaking VEV can be one: $S'$.

In the phenomenology discussion of Sec. \ref{sec:phenomenology}, the details of the ultraviolet completion does not matter because the main prodution we will consider is through the electroweak process $p\bar p\to E^-E^+$ and the $E$ decay via the interaction (\ref{eEcN}) which is common to Models {\bf a} and {\bf b}.

\section{Phenomenology}\label{sec:phenomenology}

\subsection{PAMELA high energy positrons without antiproton excess}

The model presented has the coupling (\ref{eEcN}) and explains the high energy positron excess of the PAMELA experiment \cite{TwoDMs}, but does not introduce a $u_R U^c_R N_R$ type and/or  $d_R D^c_R N_R$ type couplings with heavy quark $U$ and $D$. If such couplings were allowed, two DM annihilations would give high energy antiproton excess. But these terms cannot be present.
Firstly, these heavy quarks carrying the color charge
cannot be flipped-SU(5) singlet representations. They must appear in \ten$_1$ or \fiveb$_{-3}$ or some other SU(5) representation with a proper $X$ quantum number. Because these are heavy quarks, say $Q$ where $Q$ is $U$ or $D$, they must be vectorlike below the GUT scale, $Q_R+Q_R^c$. Both $Q$ and $Q^c$ cannot appear in \ten\ or \tenb\ because $\tent\cdot\tent\cdot N$ and $\tenbt\cdot\tenbt\cdot N$ are not allowed by the SU(5) symmetry. Similarly, both $Q$ and $Q^c$ cannot appear in \five\ or \fiveb\ because $\fivet\cdot\fivet\cdot N$ and $\fivebt\cdot\fivebt\cdot N$ are not allowed by the SU(5) symmetry. So, if one of $Q$ and $Q^c$ appears in \ten\ (or \five), the other must appear in \tenb\ (or \fiveb) to write down $Q\overline{Q}N$ coupling. But we need a coupling with a light quark $q$ where $q$ is $u$ or $d$. So, at least we must keep ten more two component fields at the electroweak scale, e.g. \five\ and \fiveb. So, assuming the minimal extension with $N$, it is just $E+E^c$. Allowing \five\ and \fiveb\ in addition will produce antiproton excess too but then we introduce ten more chiral fields. Even though we introduce \five$'$ and \fiveb$'$  and a coupling $q_R Q^c_R N_R$, deisregarding the minimality, if $m_Q$ is sufficiently heavy ($m_Q>M_\chi+m_N-m_q$) then antiproton excess is not expected. So, the flipped-SU(5) has an intrinsic mechanism to allow the positron excess without antiproton excess by an appropriate mass parameters.

Beyond this naive expectation, in fact we can show that the diagonalization of mass matrix almost forbids the antiproton excess in our flipped-SU(5) model. For the heavy vectorlike pair, we assign the quantum numbers of right-handed top for \fiveb$'$ and their opposite quantum numbers for \five$'$,
\begin{align}
&\fivebt'_{-3}:~ X=-3,~ H=1,~ R=0,~ Z_2=-,~ \Gamma=0\nonumber\\
&\fivet'_{+3}:~ X=3,~ H=-1,~ R=0,~ Z_2=-,~  \Gamma=0 ~ .\nonumber
\end{align}
The mass matrix of $Q_{\rm em}=2/3$ quarks can be written as the following matrix.
\begin{table}[!h]
\begin{tabular}{c|ll}
 &\quad $\fivebt_{-3}^{(1)}$ &\quad $\fivebt'_{-3}$\\
 \hline
  $\tent^{(1)}_{1}$\quad &
\quad $\epsilon m$&\quad $m'$\\
 $\fivet_{3}'$\quad &\quad 0 &\quad  $\tilde m$
\end{tabular}
\end{table}
Here $m$ and $m'$ are at the electroweak scale, $\tilde m$ is a free parameter, presumably around 100 GeV, and $\epsilon$ is of order $10^{-5}$ as discussed in Eq. (\ref{umass}). Then, the physical right-handed up quark  and the heavy quark are given by
\begin{align}
&u_R=\cos\theta\ u_R^{(1)} +\sin\theta~ \fivebt'_{-3}\\
&U_R=-\sin\theta~ u_R^{(1)} +\cos\theta~ \fivebt'_{-3}
\end{align}
where $\tan\theta=\epsilon m/m'$ and $\fivebt'_{-3}$ represent the $Q_{\rm em}=2/3$ quark in $\fivebt'_{-3}$. Then, the coupling $u_R U_R^c N_R$ is suppressed by $\epsilon$, and the antiproton excess is not expected in the flipped-SU(5) model.

\subsection{Production of $E$ in LHC experiments}

The model presented in Ref. \cite{TwoDMs} has three two component fermions. Among these $Q_{\rm em}=-1$ lepton $E$ can be produced at the LHC machine. The most distinguishable feature of the model presented here is leptonic (especially `electronic') property of the dark matter component. $E^+$ and $E^-$ can be produced by proton--proton collision and eventually they decay to an electron--positron pair and missing energy by the interaction (\ref{eEcN}). Then, we compare it with the expected high energy $e^+e^-$ spectrum from the LHC machine. This signal distinguishes the model of Ref. \cite{TwoDMs} from most other popular models. We will calculate the high energy $e^+e^-$ spectrum in the quark-antiquark ($q-\bar q$) center of momentum (CM) frame. Here, $q$ stands for $u,d,s,$ etc. Then, the result is convoluted to the $q$ and $\bar q$ structure functions in the proton.

For the $q-\bar q$ annihilation, the CM frame kinematics of the $q-\bar q$ pair is useful,
\begin{widetext}
\begin{eqnarray}
q+\bar q &&\longrightarrow\quad E^-(E,{\bf p})+E^+(E, -{\bf p})\\
 && \longrightarrow\quad \left\{ e^-(\epsilon_1,{\bf q}_1) +\tilde N_1(E-\epsilon_1, {\bf p}-{\bf q}_1) \right\}+\left\{ e^+(\epsilon_2,{\bf q}_2) +\tilde N_2(E-\epsilon_2, -{\bf p}-{\bf q}_2) \right\}.\nonumber
\end{eqnarray}
\end{widetext}
The $q-\bar q$ parton contribution in the proton-proton collision is given by
\begin{equation}
\sigma (pp\rightarrow e^-e^+X)= \frac{8\pi \alpha^2}{9} \sum \int dx_1 \int dx_2 f_q(x_1,Q) f_{\bar{q}}(x_2,Q) \frac{A_0(\hat{s})}{\hat{s}}.\label{cross:pp_ee}
\end{equation}
Here $f_q(x_1,Q)$ and $f_{\bar{q}}(x_2,Q)$ are parton distribution function(PDF) of $q$ and $\bar{q}$ with $p_{q}=x_1p_p$ and $p_{\bar{q}}=x_2p_p$, repectively. We use MRST99 C++ code for PDF\cite{Martin:1999ww}. Sum is for all quarks except for $t$ and is neglected from now on. $Q$ is relevent energy scale of PDF.
$\hat{s}=x_1x_2s$ with $\sqrt{s}=$14 TeV in the designed LHC energy. $A_0$ is obtained by cross section at the parton level, and is given by

\begin{equation}
A_0 = (Q_q Q_e)^2 + Q_q Q_e Re(r) (\varepsilon_L^q+\varepsilon_R^q)
(\varepsilon_L^e+\varepsilon_R^e)+|r|^2[
(\varepsilon_L^q)^2+(\varepsilon_R^q)^2][
(\varepsilon_L^e)^2+(\varepsilon_R^e)^2],
\end{equation}
where $Q_q=+2/3$ for up-type quarks $Q_q=-1/3$ for down type quarks, $Q_e=-1$ are electromagnetic charge of up and electron, and $\varepsilon_L^q=1-\frac{4}{3}\sin^2 \theta_W$, $\varepsilon_R^q=-\frac{4}{3}\sin^2 \theta_W$ for up-type quarks, $\varepsilon_L^q=-1+\frac{2}{3}\sin^2 \theta_W$, $\varepsilon_R^q=+\frac{2}{3}\sin^2 \theta_W$ for down type quarks,$\varepsilon_L^e=-1+2\sin^2 \theta_W$ and $\varepsilon_R^e=2\sin^2 \theta_W$ are $Z$ boson couplings of quarks, and electron, respectively. $\theta_W$ is the weak mixing angle \cite{KimRMP81}. $r$ is given by
\begin{equation}
r=\frac{\sqrt{2}G_FM_Z^2}{\hat{s}-M_Z^2
+iM_Z\Gamma_Z}\frac{\hat{s}}{4\pi \alpha},
\end{equation}
where $G_F$ is Fermi constant, $M_Z$ and $\Gamma_Z$ are $Z$ boson mass and decay rate.
Eq. (\ref{cross:pp_ee}) can be rewritten as
\begin{equation}
\begin{split}
\sigma (pp\rightarrow e^-e^+X) &= C\int d\hat{s} \int dx_1 \int dx_2 \delta(\hat{s}-x_1x_2s) f_q(x_1,Q) f_{\bar{q}}(x_2,Q) \frac{A_0(\hat{s})}{\hat{s}}\\
&=C\int d\hat{s} \int dx_1 \int dx_2 \frac{1}{x_1s}\delta(x_2-\frac{\hat{s}}{x_1s})f_q(x_1,Q) f_{\bar{q}}(x_2,Q) \frac{A_0(\hat{s})}{\hat{s}}\\
&=C\int d\hat{s} \int dx_1 \frac{1}{x_1s}f_q(x_1,Q) f_{\bar{q}}(\frac{\hat{s}}{x_1s},Q) \frac{A_0(\hat{s})}{\hat{s}}\\
&=\int d\hat{s} \frac{d\sigma}{d\hat{s}}(pp\rightarrow e^-e^+X)
\end{split}
\end{equation}
where
\begin{equation}
\frac{d\sigma}{d\hat{s}}(pp\rightarrow e^-e^+X) = C\int^1_{\hat{s}/s} dx_1 \frac{1}{x_1s}f_q(x_1,Q) f_{\bar{q}}(\frac{\hat{s}}{x_1s},Q) \frac{A_0(\hat{s})}{\hat{s}},
\end{equation}
which is cross section between $\hat{s}+d\hat{s}$ and $\hat{s}$. Here $C=8\pi \alpha^2 /9$.

\begin{figure}[!]
\vskip 0.5cm
\resizebox{0.9\columnwidth}{!}
{\includegraphics{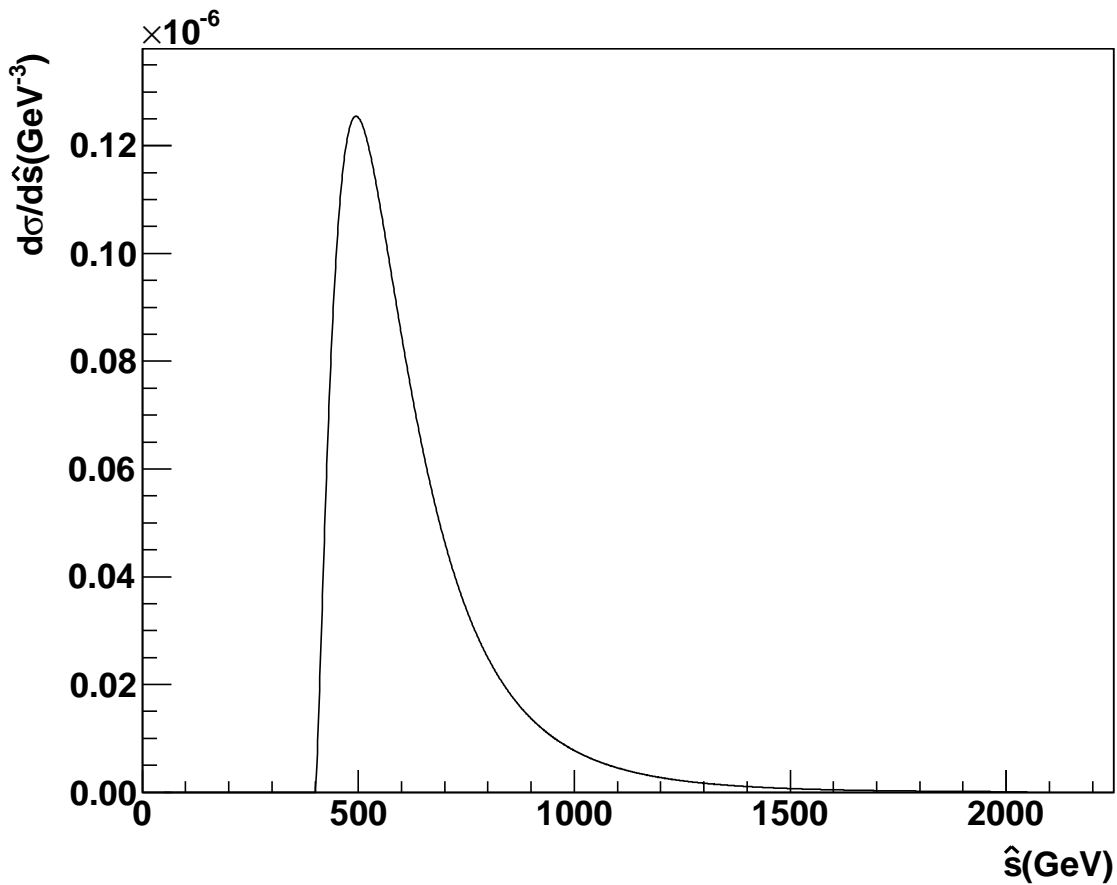}}
\caption{The $E^+E^-$ production cross section. The horizontal axis is the CM energy in GeV.  The unit of the vertical axis is GeV$^{-2}$.}\label{fig:EEprod}
\end{figure}

Similarly, $\frac{d\sigma}{d\hat{s}}(pp\rightarrow E^-E^+X)$ is obtained by simple calculations, which is given by
\begin{equation}
\frac{d\sigma}{d\hat{s}}(pp\rightarrow E^-E^+X) = \frac{8\pi \alpha^2}{9} \int^1_{\hat{s}/s} dx_1 \frac{1}{x_1s}f_q(x_1,Q) f_{\bar{q}}(\frac{\hat{s}}{x_1s},Q) \frac{A^{\prime}_0(\hat{s})}{\hat{s}}\biggl( 1-\frac{1}{\gamma^2}\biggr)^{3/2},
\end{equation}
where $\gamma=E_E/m_E$. $A_0^{\prime}$ is given by
\begin{equation}
A_0^{\prime} = (Q_q Q_E)^2 + Q_q Q_E Re(r) (\varepsilon_L^q+\varepsilon_R^q)
(\varepsilon_L^E+\varepsilon_R^E)+|r|^2
[(\varepsilon_L^q)^2+(\varepsilon_R^q)^2]
[(\varepsilon_L^E)^2+(\varepsilon_R^E)^2],
\end{equation}
where $Q_E=-1$ and $\varepsilon_L^E=\varepsilon_R^E=2\sin^2\theta_W$.
The production cross section of $E^-E^+$ is approximately a half of that of electron-positron pair at weak scale. Cross section is shown in Fig. \ref{fig:EEprod}.

\begin{figure}[!]
\vskip 0.5cm
\resizebox{0.9\columnwidth}{!}
{\includegraphics{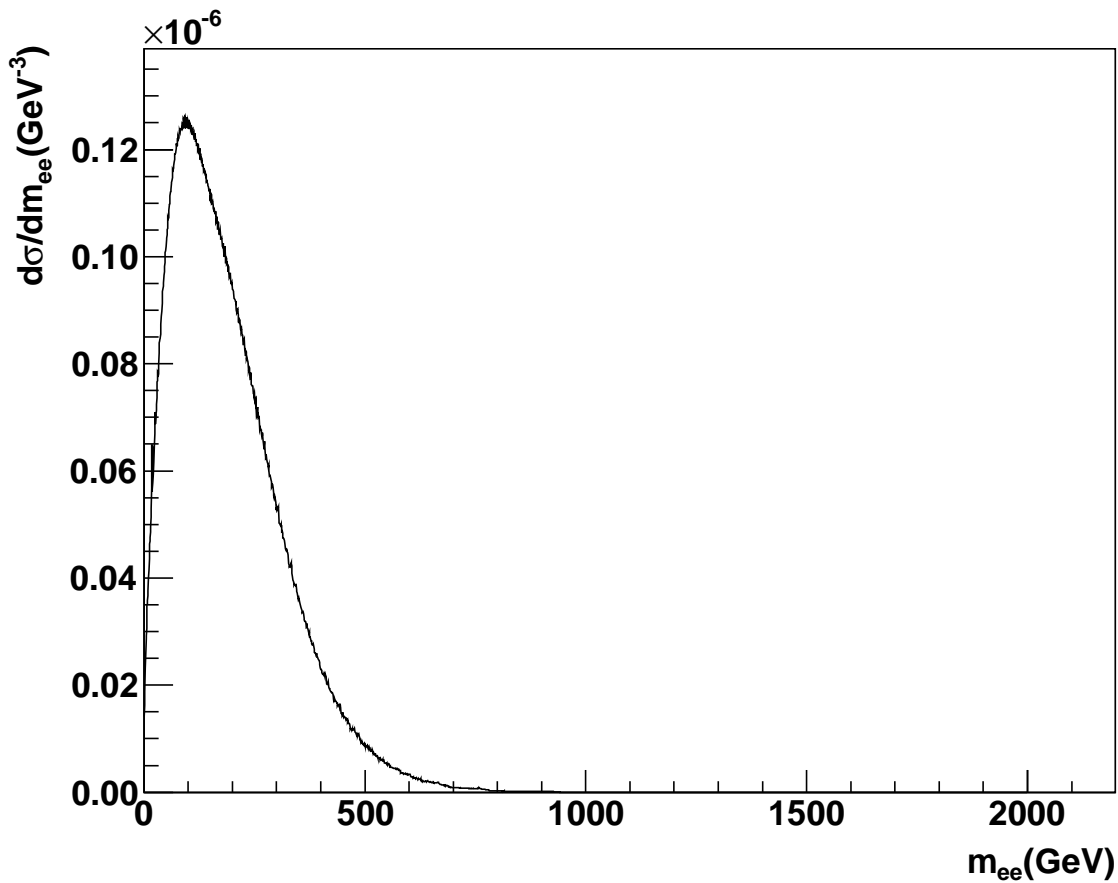}}
\caption{The expected $e^+e^-$ production through $E^+E^-$ decay. The horizontal axis is the invariant pass of $e^+e^-$ pair in GeV. The unit of the vertical axis is GeV$^{-2}$.}\label{fig:eefromEE}
\end{figure}

However, since $E$ decays inside the beam line, electron-positron excess appears at lower energy scale. Therefore we can detect electron-positron pairs in collider. Cross sectrion can be rewritten in elctron-positron transverse invariant mass $m_{ee}\equiv \sqrt{(q_1+q_2)^2}=\sqrt{\epsilon_{1T}\epsilon_{2T}
\cosh(\Delta\eta)-\bold{q}_{1T}\cdot\bold{q}_{2T}}$ where $\epsilon_{iT}$, $\bold{q}_{iT}$ and $\Delta \eta$ are transverse energy, transverse momentum and rapidity difference, respectively. It is given by
\begin{equation}
\sigma(pp\rightarrow E^-E^+X \rightarrow e^-e^+X) = \int dm_{ee} \frac{d\sigma}{dm_{ee}} (pp\rightarrow E^-E^+X \rightarrow e^-e^+X)
\end{equation}
where
\begin{equation}
\begin{split}
\frac{d\sigma}{dm_{ee}} (pp\rightarrow E^-E^+X \rightarrow e^-e^+X) &= \sigma(pp\rightarrow E^-E^+X)\\
&\times\int d\Omega_- d\Omega_+ \frac{d}{dm_{ee}} \biggl[ \frac{1}{\Gamma_-}\frac{d\Gamma_-}{
d\Omega_-}(E^-\rightarrow e^-+\tilde{N}) \frac{1}{\Gamma_+}\frac{d\Gamma_+}{d\Omega_+}
(E^+\rightarrow e^++\tilde{N})\biggr]
\end{split}
\end{equation}
Also, these events are accompanied by large missing energy and this physics is very similar to slepton pair production of odinary MSSM. This is shown in Fig. \ref{fig:eefromEE}.

The main difference of our calculation from the ordinary MSSM result is the produced $E$'s totaly decay to electrons but not to the other leptons. Due to this feature, we can use transverse momentum variable, $m_{T2}$ to determine the mass of $E$ \cite{Lester99}. Note that the phenomenology of $E^+E^-$ production is very similar to that of the selectron--antiselectron production because in both cases the heavy particle decays to $e^\pm$ plus missing energy. So, it is crucial to distinguish them. But, the only way to distinguish them is to measure their spin at the LHC.
Below, we concentrate on the total cross section but not the angular distribution. In fact, there are many analyses for measuring the particle spin at the LHC machine \cite{Datta:2005zs,Choi:2006mr}. Since our $E$ particle is the fermion, their polarization property looks the same as the KK fermion in the UED models, and it may be difficult to distinguish our fermion from the UED fermions. But we can expect to distinguish $E$ from the selectron. For the same $E$ and selectron mass, their kinematics are similar but angular distribution of electrons might be quite different in the two cases. If both $E^+E^-$ and the selectron--antiselectron pair are produced, we can observe two $m_{T2}$s which might be spectacular.

\section{Conclusion}\label{sec:conclusion}

The recent PAMELA satellite data from the galactic source suggested the positron excess \cite{PAMELAe} but no anti-proton excess \cite{PAMELAp}. Regarding these observations, two dark matter components, the neutralino $\chi$ and the singlet fermion $N$, were suggested in Ref. \cite{TwoDMs}. The needed coupling has been given by Eq. (\ref{eEcN}). Here, we studied the flipped-SU(5) where charged SU(2) singlet leptons such as $e_R, E_R$ and $E_R^c$ of  Eq. (\ref{eEcN}) can be its allowed representations, but a similar coupling for an anti-proton excess is not allowed because quarks must be embedded in SU(5) non-singlet representations.

In addition, there exists an interesting suggestion of the possible existence of VWLP with the coupling strength (\ref{aeWDcoupling}) from the study of white dwarf cooling \cite{Isern08}. In this paper, we tried to interpret this VWLP as an  {\it electrophilic} axion with the axion decay constant $F_a\simeq 1.4\times 10^{10}$ GeV. In the DFSZ model, the corresponding $F_a$ would be $\simeq 1.2\times 10^{9}$ GeV which would be barely consistent with the SN1987A bound. This kind of  {\it electrophilic} axion can arise in `variant very light axion' models where different families possess different PQ quantum numbers $\Gamma$. We argued that the inverted mass pattern of the first family from that of the second and third families is the logic for assigning different PQ quantum numbers for different family members. Allowing an effective domain wall number of $\frac12$ [See Footnote \ref{footDW}.], we constructed an {\it electrophilic} axion.
The symmetries we introduced in (\ref{symmetry}) is bigger than \suf1$\times$U(1)$_\Gamma$, to allow a GUT scale VEV with an intermediate scale $F_a$ and R-parity.
With these symmetries, we explained the mass hierarchy between the top and bottom quarks. In this model, the gauge couplings and the top quark Yukawa coupling are the only renormalizable couplings.

This model needs a charged SU(2) singlet lepton $E$ at the electroweak scale. So, the existence of $E$ is absolutely needed for this suggestion of the positron excess to work. Here, the discovery potential of $E$, at the LHC experiments by observing the electron and positron spectrum, is presented. We calculated the $e^+e^-$ spectrum at the center of momentum frame of the quark and anti-quark pair and compared it with the expected background. We pointed out that the SU(2) singlet $E$ with the coupling (\ref{eEcN}) is distinguishable from other possibilities we can imagine. Therefore, the discovery of $E$ at the LHC experiments will indirectly confirm the two dark matter component hypothesis of \cite{TwoDMs}. Or it can be ruled out from the LHC experiments by the absence of the $e^+e^-$ excess above the background.

\vskip 0.3cm
\acknowledgments{R.D.V. thanks the discussion with Brian Warner. K.J.B, J.H.H., and J.E.K. thanks the Dept. of Physics of the University of Cape Town for the hospitality extended to them while this work was finished.
K.J.B, J.H.H., J.E.K. and B.K are supported in part by the Korea Research Foundation, Grant No. KRF-2005-084-C00001 of Ministry of Education, Science and Technology(MEST) of Republic of Korea. In addition, K.J.B is supported in part by the Korea Science and Engineering Foundation grant funded by the MEST through the Center for Quantum Spacetime of Sogang University with Grant Number R11-2005-021, B.K. is supported by the FPRD of the BK21 program and the KICOS Grant No. K20732000011-07A0700-01110 of the MEST. R.D.V. is supported in part by  the Foundation for the Fundamental Research (FFR) Grant Number PHY99-1241, the National Research Foundation of South Africa Grant No. FA2005033 100013, and the Research Committee of the University of Cape Town.
}



\end{document}